\documentclass[onecolumn]{emulateapj-rtx4}
\usepackage{longtable}
\usepackage{graphicx}
\usepackage{graphics}
\usepackage{color}
\usepackage{epsfig}
\usepackage{lscape}

\shortauthors{}
\shorttitle{}
\begin{document}

\title{Dynamic S0 Galaxies II: the Role of Diffuse Hot Gas}

\author{Jiang-Tao Li\altaffilmark{1, 2},
Q. Daniel Wang\altaffilmark{2}, Zhiyuan Li\altaffilmark{3} and Yang
Chen\altaffilmark{1}} \altaffiltext{1}{Department of Astronomy,
Nanjing University, 22 Hankou Road, Nanjing 210093, P. R. China}
\altaffiltext{2}{Department of Astronomy, University of
Massachusetts, 710 North Pleasant Street, Amherst, MA 01003, U.S.A.}
\altaffiltext{3}{Harvard-Smithsonian Center for Astrophysics, 60
Garden Street, Cambridge, MA 02138}


\begin{abstract}
Cold gas loss is thought to be important in star formation quenching and morphological transition during the evolution of S0 galaxies. In high density environments, this gas loss can be achieved via many external mechanisms. However, in relatively isolated environments, where these external mechanisms cannot be efficient, the gas loss must then be dominated by some internal processes. We have performed \emph{Chandra} analysis of hot gas in five nearby isolated S0 galaxies, based on the quantitative subtraction of various stellar contributions. We find that all the galaxies studied in the present work are X-ray faint, with the luminosity of the hot gas ($L_X$) typically accounting for $\lesssim5\%$ of the expected Type~Ia SN energy injection rate.
We have further compared our results with those from relevant recent papers, in order to investigate the energy budget, cold-hot gas relation, and gas removal from S0 galaxies in isolated environments. We find that elliptical and S0 galaxies are not significantly different in $L_X$ at the low mass end (typically with K-band luminosity $L_K\lesssim10^{11}\rm~L_{\odot,K}$). However, at the high mass end, S0 galaxies tend to have significantly lower $L_X$ than elliptical galaxies of the same stellar masses, as already shown in previous observational and theoretical works. We further discuss the potential relationship of the diffuse X-ray emission with the cold (atomic and molecular) gas content in the S0 and elliptical galaxies included in our study. We find that $L_X/L_K^2$ tends to correlate positively with the total cold gas mass ($M_{H_2+HI}$) for cold-gas-poor galaxies with $M_{H_2+HI}\lesssim10^8\rm~M_\odot$, while they anti-correlate with each other for cold-gas-rich galaxies. This cold-hot gas relationship can be explained in a scenario of early-type galaxy evolution, with the leftover cold gas from the precursor star forming galaxy mainly removed by the long-lasting Type~Ia SN feedback. The two different trends for cold-gas-rich and -poor galaxies may be the results of the initial fast decreasing SN rate and the later fast decreasing mass-loading to hot gas, respectively.
\end{abstract}

\section{Introduction}\label{sec:Introduction}

Lenticular, or S0 galaxies, sitting on the intersection of the Hubble sequence, are often thought to be the remnants of spiral galaxies after their star formation (SF) has ceased or slowed down substantially (Bekki et al. 2002; van den Bergh 2009a, b). This reduction of the SF rate (SFR) must be directly related to the decrease of the cold gas content and/or density in the galactic disk. The change could be caused by various internal or external mechanisms (e.g., Martig et al. 2009 and references therein), such as ram-pressure stripping (of cold gas), termination or strangulation (removing surrounding hot gas, the cooling of which may be a source of new cold gas), morphological quenching (heating of a stellar disk or even transforming to a spheroid), and exhaustion (consuming gas via SF). The relative importance of these mechanisms cannot be easily determined because large uncertainties remain in our understanding of these processes, especially in terms of their connection to the observed properties of the interstellar medium (ISM). However, if only galaxies in relative isolation are considered, where the external mechanisms are likely ineffective, the removal of the cold gas must then be dominated by the internal processes.

Type~Ia SN feedback is one such internal process, especially in S0 galaxies with a modest mass and a low SFR. Li et al. (2009) (paper I) found extraplanar filamentary dusty features in the edge-on S0 galaxy NGC~5866, which they suggested is cold gas blown out from the galactic disk. For individual dusty filaments, the energy required to lift them is comparable to the energy of a single SN. The SF in this galaxy is inactive and less energetic than Type~Ia SNe. The latter is thus likely the origin of these features and the dominant source of cold gas removal in this special case.

Currently there is still not much direct evidence for gas removal by Type~Ia SNe in S0 galaxies. However, cold gas is detected in many of them, either produced by evolved star mass loss or leftover from precursor spiral galaxies (Sage \& Welch 2006). Possible interactions between the cold gas and the hot gas produced by SN heating may have various forms as: photo evaporation (Melioli et al. 2005), turbulent mixing due to Kelvin-Helmholtz instability (Cooper et al. 2009; Pittard et al. 2010), and enhanced radiative cooling (Melioli \& de Gouveia dal Pino 2004; Melioli et al. 2005; Cooper et al. 2009), etc. These interactions can have significant effects on the observed hot gas properties. In particular, the mass loading of cold gas into the hot phase can increase the density, while reduce the specific energy or temperature of the hot gas, so enhance the observed soft X-ray emission. In addition, the reduced specific energy makes the hot gas take more time to flow out and better mix with the interstellar medium (ISM), resulting in a lower SN heating efficiency, i.e., the fraction of the SN energy used to drive the outflow (Melioli \& de Gouveia dal Pino 2004; Strickland \& Heckman 2009). The dynamical state and the observed soft X-ray luminosity of the hot gas thus crucially depend on the cold gas content; the cold-hot gas interaction behaves like a glue that helps to make the gas easier to be gravitationally bound.

Although indirect evidence of this cold-hot gas interaction, such as the complicated thermal structure of the hot gas, has been revealed in some late-type galaxies (e.g., Strickland et al. 2004; Li et al. 2008; Owen \& Warwick 2009), it is difficult to find direct evidence for such interaction in early-type galaxies. This is because early-type galaxies often have a low cold gas content and their hot gas luminosity is usually thought to be determined by stellar mass instead of the cold gas properties. However, the relations between the hot gas luminosity ($L_X$) and various stellar mass tracers, such as B-band or K-band luminosities ($L_B$ or $L_K$), typically show scatters in $L_X$ as large as $\gtrsim3$ orders of magnitude at a certain $L_B$ or $L_K$ value (Mathews \& Brighenti 2003). Many possibilities have been discussed to explain such a large scatter (e.g., Ellis \& O'Sullivan 2006; for a review, refer to Mathews \& Brighenti 2003), including AGN heating (e.g., David et al. 2006), environmental effects (e.g., White \& Sarazin 1991; Helsdon et al. 2001; Mulchaey \& Jeltema 2010), various dynamical states of gas flows (e.g., Ciotti et al. 1991), different stellar mass distributions (e.g., Brighenti \& Mathews 1996; Pellegrini 1999), different ages of the stellar populations (e.g., Memola et al. 2009; Boroson, Kim \& Fabbiano 2010), and various dark matter halo to stellar mass ratios (e.g., Mathews et al. 2006), etc. In addition to these mechanisms, due to the cold-hot gas interaction described above, it is also possible that the various cold gas contents in early-type galaxies may make some contributions to the scatter of this stellar mass-$L_X$ relation.

In this paper, we present \emph{Chandra} observations of five nearby isolated S0 galaxies to study the hot gas properties of them, and explore the relations between the cold/hot gases. The paper is organized as follows: Sample selection is presented in \S\ref{sec:SampleSelection}. Detailed data reduction is described in \S\ref{sec:Data}. We present our results on individual galaxies in \S\ref{sec:results}. We further present some statistical analysis and discuss their implications in \S\ref{sec:Discussion}. Finally we summarize our main results in \S\ref{sec:Summary}. Errors of the data obtained in this work are quoted at the 90\% confidence level.

\section{Sample Selection}\label{sec:SampleSelection}

Analysis of the \emph{Chandra} data of many early-type galaxies are already available in literatures (e.g., Mulchaey \& Jeltema 2010; Boroson et al. 2011). In this paper, we concentrate on nearby isolated S0 galaxies, aiming at comparing their hot gas properties to those of elliptical galaxies and studying the cold/hot gas relation. Our sample selection adopts the following criteria:

$\bullet$ \emph{Morphological classification.} We select galaxies which are optically classified as E-S0, S0, or S0-a in the \emph{HyperLeda} database, or with morphological type code $-3\lesssim TC\lesssim0.5$.

$\bullet$ \emph{Distance.} We select only galaxies with a distance $d\lesssim20\rm~Mpc$, so that we can resolve the bulk of the luminous point sources. The distances to NGC~2681, NGC~3115, and NGC~5866 are measured from the I-band surface brightness fluctuation (SBF; Tonry et al. 2001). For other galaxies, either because the SBF measurement is not available (NGC~1291) or the galaxy has too complicated bulge structure for such a distance measurement (NGC~2787), distances are estimated from the receding velocity, adopting the local velocity field model given in Mould et al.~(2000) (from the \emph{NED} database).

$\bullet$ \emph{Environment.} Focusing on the internal processes, we select relatively isolated galaxies, which have local galaxy number density $\rho\lesssim 0.6$ (Table~\ref{table:basicpara}). None of the galaxies have close companions, which may tidally affect the gas content and dynamics significantly.

$\bullet$ \emph{Foreground extinction.} Because hot gas emission in nearby galaxies peaks in soft X-rays, our sample galaxies all have Galactic foreground \ion{H}{1} column density $N_H\lesssim5\times10^{20}\rm~cm^{-2}$ to minimize the extinction.

$\bullet$ \emph{Data requirement.} All the selected galaxies have archival \emph{Chandra} data with a total exposure time $t_{exp}\gtrsim30\rm~ks$.

As a result, only five galaxies are selected. Logs of galaxy parameters and the \emph{Chandra} observations are listed in Table~\ref{table:basicpara} and columns (1-3) of Table~\ref{table:hotgas} respectively.

\begin{deluxetable}{ccccccccccccccc}
\centering
\scriptsize 
  \tabletypesize{\scriptsize}
  \tablecaption{Basic Parameters of the Sample Galaxies}
  \tablewidth{0pt}
  \tablehead{\colhead{Name} & \colhead{Type} & \colhead{$TC$} & \colhead{$d$} & \colhead{$N_H$} & \colhead{$L_K$} & \colhead{$M_\ast$} & \colhead{$\rho$} & \colhead{$M_{HI}$} & \colhead{$M_{H_2}$} \\
 (NGC)  &  &  & (Mpc) & ($10^{20}\rm cm^{-2}$) & ($10^{10}\rm L_{\odot,K}$) & ($10^{10}\rm M_{\odot}$) & ($\rm Mpc^{-3}$) & ($10^{8}\rm M_{\odot}$) & ($10^{8}\rm M_{\odot}$)\\
   & (1) & (2) & (3) & (4) & (5) & (6) & (7) & (8) & (9)
}
  \startdata
 1291 & S0-a & $0.1\pm0.4$  & $10.1\pm0.7$ & 2.12 & 10.2& 8.0 & 0.14 & 8.1 & 0.22 \\
 2681 & S0-a & $0.4\pm0.6$  & $17.2\pm3.7$ & 2.45 & 6.2 & 4.0 & 0.20 & $<0.095$ & 0.19 \\
 2787 & S0-a & $-1.1\pm0.7$ & $14.3\pm1.0$ & 4.32 & 4.9 & 3.8 & 0.06 & 7.68 & 0.178 \\
 3115 & E-S0 & $-2.9\pm0.5$ & $9.7\pm0.5$  & 4.32 & 7.0 & 5.5 & 0.08 & $<0.0255$ & $<0.0333$ \\
 5866 & S0-a & $-1.3\pm0.7$ & $15.3\pm1.1$ & 1.46 & 8.8 & 5.8 & 0.24 & 1.20 & 4.39
\enddata
\tablecomments{\scriptsize Basic parameters of the sample galaxies. The morphological type (1) and type code (2) are obtained from the \emph{HyperLeda} database (\emph{http://leda.univ-lyon1.fr/}). The foreground HI column density (4) is obtained from the HEASARC web tools (\emph{http://heasarc.gsfc.nasa.gov/docs/tools.html}). For NGC~1291 and NGC~2787, distances (3) are obtained based on the local velocity field model given in Mould et al.~(2000) using the terms for the influence of the Virgo Cluster, the Great Attractor, and the Shapley Supercluster (from the \emph{NED} database: \emph{http://nedwww.ipac.caltech.edu/}). For other galaxies, distances are obtained from the I-band surface brightness fluctuation (SBF) measurements (Tonry et al. 2001). The K-band luminosity (5) and the stellar mass (6) are obtained from the present work (\S\ref{sec:Data}). The density of galaxies brighter than -16~mag in the vicinity of the galaxy (7) is obtained from the Nearby Galaxies Catalog (Tully 1988). The masses of atomic (8) and molecular (9) gases of most of the galaxies are obtained from Welch \& Sage (2003), Sage \& Welch (2006), and Welch et al. (2010), while those for NGC~2681 are obtained from Haynes et al. (1988) and Taniguchi et al. (1994).}\label{table:basicpara}
\end{deluxetable}

\section{Data Reduction and Analysis}\label{sec:Data}

We reprocess the data using CIAO v.4.1 and the corresponding calibration files, following \emph{Chandra} data analysis guide. We extract background lightcurves and remove obvious background flares, resulting in the effective exposure time $t_{eff}$ in column (3) of Table~\ref{table:hotgas}. We perform source detection in the broad (B, $0.3-7\rm~keV$), soft (S, $0.3-1.5\rm~keV$) and hard (H, $1.5-7\rm~keV$) bands, following the procedure detailed in Wang (2004). To study the diffuse X-ray emission, we remove the detected discrete sources from the data. Circular regions are excluded within twice the 90\% energy enclosed radius (EER) around each source of a count rate $(CR)\lesssim 0.01\rm~cts~s^{-1}$. For brighter sources, the removal radius is further multiplied by a factor of $1+log(CR/0.01)$. Generally about 96\% of the source counts are excluded in such a removal.

After removing these relatively bright point-like sources, the unresolved emission in the soft band mainly consists of hot gas plus various contributions from faint stellar X-ray sources (typically with $L_X\lesssim10^{36-37}\rm~erg~s^{-1}$), such as low luminosity low mass X-ray binaries (LMXBs), cataclysmic variables (CVs), and coronally active binaries (ABs). Recently, Revnivtsev et al. (2007, 2008, 2009) have calibrated the X-ray emissivity of CVs and ABs in the Galactic ridge as well as in some nearby low-mass early-type galaxies which are extremely gas-poor. These calibrations allow us to isolate the stellar contribution and further quantify the emission from diffuse hot gas.

We analyze the spectra of the unresolved X-ray emission following the steps detailed in Paper~I. The spectra are extracted from the inner bulge to optimize the signal-to-noise ratio. They are then subtracted with the local sky background and modeled with a thermal plasma (\emph{MEKAL} or \emph{VMEKAL} in XSPEC) plus various stellar contributions: (1) the emission from CVs and ABs, (2) the emission from unresolved LMXBs below the detection threshold, as well as residual photons after the source removal. The second term is referred to as ``the unresolved LMXB component'' hereafter. We quantify the CV+AB contribution using the 0.5-2~keV emissivity (per unit stellar mass) of $\sim(7.0\pm2.9)\times10^{27}\rm~ergs~s^{-1}~M_\odot^{-1}$, given by Revnivtsev et al.~(2008). The spectrum of this CV+AB contribution is characterized by a model consisting of a \emph{MEKAL} with $kT=0.5\rm~keV$ and a power law with a photon-index $\Gamma=1.9$ (Revnivtsev et al. 2008). The unresolved LMXB component is modeled by a power law with the same photon index as the accumulated discrete stellar X-ray source spectrum (the luminosity of this component, $L_{fit}$, is listed in Table~\ref{table:hotgas}). We further estimate the contribution of this unresolved LMXB component from the expected fraction of the spilling out photons ($\sim4\%$) and the LMXB luminosity function (LF) from Gilfanov (2004) (the luminosity estimated this way, $L_{predict}$, is also listed in Table~\ref{table:hotgas}). Such an estimation is further compared to the above fitting result to verify the validity of the spectral modeling. Generally, the agreement between $L_{fit}$ and $L_{predict}$ is within 50\% (Table~\ref{table:hotgas}). For ease of comparison with literature results, we further renormalize the X-ray luminosity of hot gas to an ellipse with the major axis of D25 and an axis ratio obtained from the optical isophote (from \emph{HyperLeda}). The renormalization factor is determined by normalizing the net diffuse soft X-ray flux (0.5-1.5~keV, instrumental-background-subtracted and exposure-corrected) in this elliptical region to that in the spectral analysis region. Results of the spectral analysis are summarized in Table~\ref{table:hotgas}.

\begin{deluxetable}{cccccccccccccc}
\centering
\tiny 
  \tabletypesize{\tiny}
  \tablecaption{\emph{Chandra} Data and Spectral Analysis Results}
  \tablewidth{0pt}
  \tablehead{
\colhead{Name} & \colhead{ObsID} & \colhead{$t_{exp}$} & \colhead{$t_{eff}$} & \colhead{$L_{X,spec}$} & \colhead{$L_{X,cor}$} & \colhead{$T_X$} & \colhead{$L_{CV,AB}$} & \colhead{$L_{fit}$} & \colhead{$L_{predict}$} \\
  & (1) & (2) & (3) & (4) & (5) & (6) & (7) & (8) & (9)
} \startdata
1291 &795+2059 & 76.7 & 58.4 & $17.1\pm1.1$ & $30.2_{-2.0}^{+1.9}$ & $0.34\pm0.04$  & 3.9 & $<$3.0 & 1.9 \\
2681 &2060+2061& 161.9 & 149.8 & $17.1\pm1.3$ & $18.0\pm1.4$ & $0.25_{-0.02}^{+0.04}$ & 2.3 & 3.4 & 2.3 \\
2787 & 4689 & 31.2 & 30.2 & $0.7_{-0.5}^{+0.3}$ & $0.7_{-0.5}^{+0.3}$&$0.15_{-0.15}^{+0.05}$ & 1.7 & 3.2 & 3.3 \\
3115 & 2040 & 37.5 & 35.1 & $1.0_{-0.8}^{+1.2}$& $1.9_{-1.5}^{+2.2}$ & $0.31_{-0.12}^{+1.10}$ & 1.8 & 2.3 & 1.5 \\
5866 & 2879 & 34.2 & 27.8 & $10.6_{-2.3}^{+2.0}$& $11.5_{-2.5}^{+2.1}$ & $0.14_{-0.001}^{+0.004}$& 2.6 & 7.8 & 5.0
\enddata
\tablecomments{\scriptsize The parameters inferred from the calibration and analysis of the X-ray data. All the luminosities are in 0.5-2~keV band in unit of $10^{38}\rm~ergs~s^{-1}$. (1) The \emph{Chandra} observation ID of the data used in the present work. (2) The total exposure time in kilo-second. (3) The effective exposure time after background flare removal. (4) Hot gas luminosity in the spectral analysis region. (5) Hot gas luminosity renormalized to an elliptical region as described in \S\ref{sec:Data}. (6) Temperature of the hot gas in keV. (7) Luminosity of the CV/AB component in the spectral analysis region. (8) Luminosity of the unresolved LMXB component measured from spectral analysis. (9) Luminosity of the unresolved LMXB component in the spectral analysis region, as predicted using the stellar mass, the LMXB luminosity function, and the accumulated stellar X-ray source spectrum. The quoted errors in columns (4-6) are only the fitting error. Uncertainties in data calibration and CV/AB subtraction are not included. Refer to \S\ref{sec:Data} for details.}\label{table:hotgas}
\end{deluxetable}

We use \emph{Spitzer} and \emph{2MASS} data to trace the dust and stellar emissions from these S0 galaxies. We adopt the aperture correction factors from Dale et al. (2007) for photometry of the \emph{Spitzer} IRAC and MIPS images. The stellar contribution is further subtracted from the mid-IR images (\emph{Spitzer} MIPS 24~$\rm\mu m$ and IRAC 8~$\rm\mu m$) using the near-IR emission (IRAC 3.6~$\rm\mu m$) and Hunter et al. (2006)'s normalization factors. The 24~$\rm\mu m$ and 8~$\rm\mu m$ images so obtained are dominated by emissions from dust grain and PAH respectively. These calibrated mid- and near-IR images are shown in Fig.~\ref{fig:image}. We adopt a color-dependent stellar mass-to-light ratio from Bell \& de~Jong (2001) when estimating the stellar mass of the whole galaxy using the \emph{2MASS} (Two Micron All Sky Survey) K-band luminosity (Table~\ref{table:basicpara}). The K-band photometry region is the same as the ellipse used in renormalizing the X-ray luminosity, with bright nuclear and foreground sources subtracted.

\section{Results}\label{sec:results}

Based on the decomposition of the unresolved soft X-ray emission (\S\ref{sec:Data}), we find that much of it is stellar in origin. The emission from the unresolved stellar X-ray sources (CV, AB, and faint LMXBs) is usually comparable to or even stronger than that from the truly diffuse hot gas (Table~\ref{table:hotgas}). After accounting for the uncertainty in the CV+AB ($<$40\% in 0.5-2~keV range; Revnivtsev et al. 2008; Boroson et al. 2011) and unresolved LMXB (typically $<$50\%, comparing the fitting results with those predicted by the LF) contributions, three of our sample galaxies (NGC~1291, NGC~2681, and NGC~5866) still show significant amounts of truly diffuse hot gas. However, NGC~2787 and NGC~3115 only have a marginal detection of the hot gas emission (Table~\ref{table:hotgas}). The soft X-ray luminosities of these X-ray faint galaxies are significantly lower than those presented in previous works without the CV+AB contribution subtracted (e.g., David et al. 2006). In addition, the subtraction of the stellar contribution, which has a relatively hard spectrum, results in a lower temperature of the hot gas (e.g., compared to the results in David et al. 2006 for NGC~2787 and NGC~5866; the error for NGC~3115 is too large for a reasonable comparison).

\begin{figure}[!h]
\begin{center}
\epsfig{figure=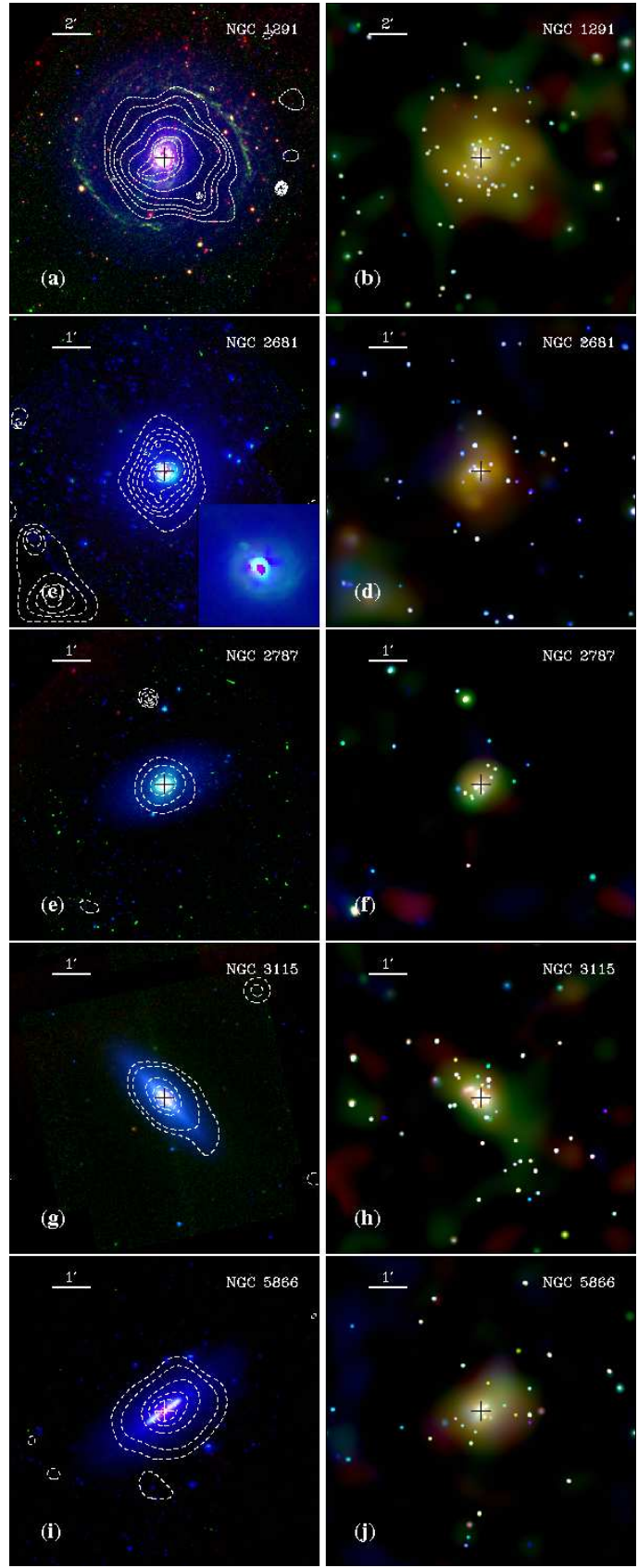,width=0.5\textwidth,angle=0, clip=}
\caption{Tri-color images of the sample galaxies. The pluses mark the optical center of the galaxies. For panels in the \emph{left} column, {\sl Red}: {\sl Spitzer} MIPS channel 1 image ($24~\rm\mu m$); {\sl Green}: \emph{Spitzer} IRAC channel 4 image ($8~\mu m$) with stellar contribution subtracted (the residual emission is mainly from PAH); {\sl Blue}: {\sl Spitzer} IRAC channel 1 image ($3.6~\mu m$). For NGC~2787 and NGC~3115, the stellar contribution is not subtracted from the $8~\mu m$ image, as the truly PAH emission is too weak. The contours are the smoothed discrete-source-removed \emph{Chandra} 0.5-1.5~keV intensities. The small box in the lower right corner of panel (c) is a zoom-in image of the central $1'\times1'$ of NGC~2681 to better show the nuclear structures. Panels in the \emph{right} column show the smoothed \emph{Chandra} images, {\sl Red}: 0.5-0.8 keV, {\sl Green}: 0.8-1.5 keV, {\sl Blue}: 1.5-7 keV.}\label{fig:image}
\end{center}
\end{figure}

The mid-IR images do reveal some dusty features (Fig.~\ref{fig:image}), which suggest the existence of cold gas in these S0 galaxies. In the following, we describe specific characteristics of individual galaxies, focusing on the observed cold/hot gas features.

\begin{figure}[!h]
\begin{center}
\epsfig{figure=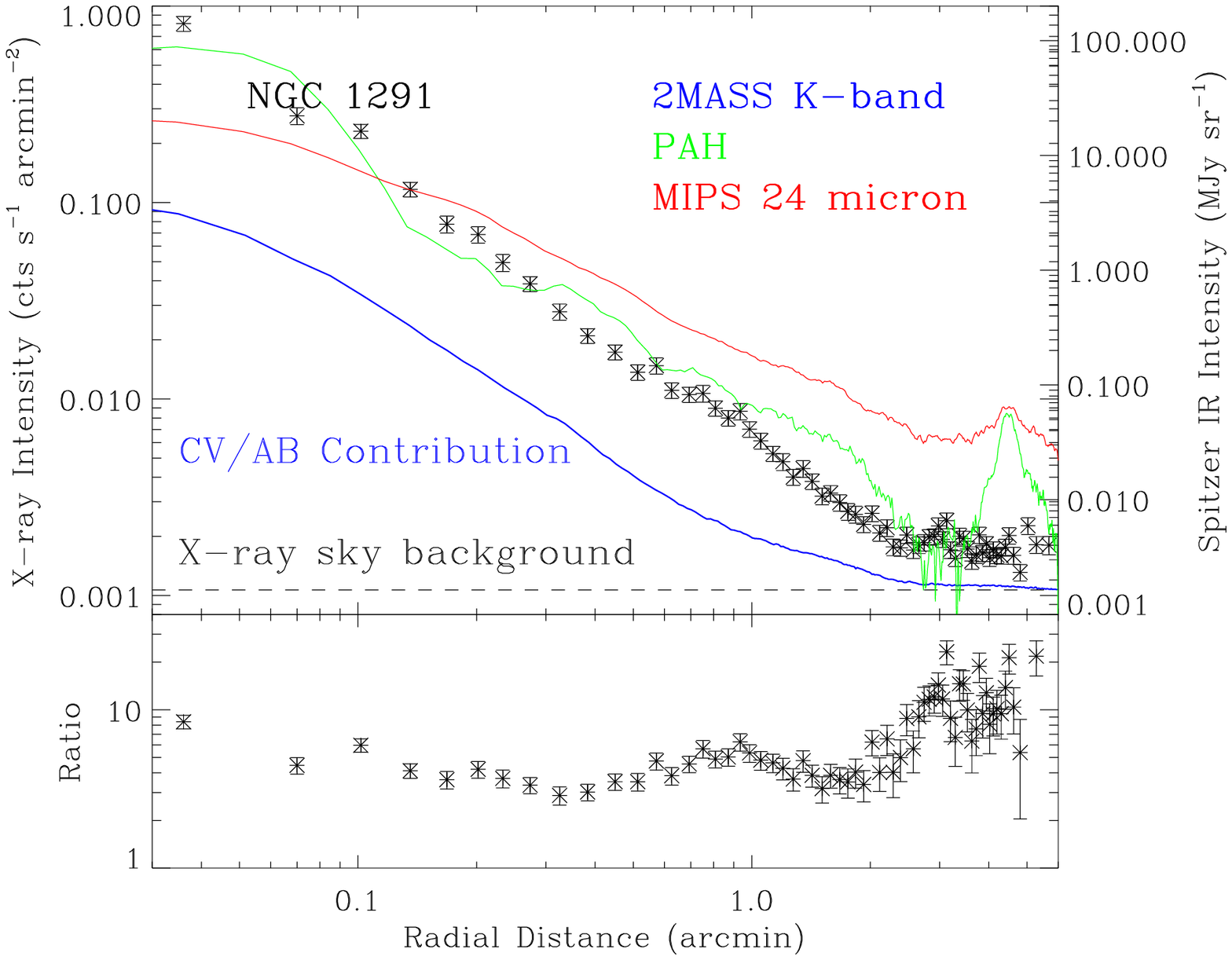,width=0.7\textwidth,angle=0, clip=}
\caption{Multi-wavelength radial brightness distributions of NGC~1291. \emph{Upper panel:} The soft X-ray (black data points) and IR (colored curves) profiles. The 2MASS K-band profile is normalized to represent the CV/AB contribution to the soft X-ray band, using the model described in \S\ref{sec:Data}. \emph{Lower panel:} The ratio between the hot gas emission (total soft X-ray emission minus the CV/AB contribution, the unresolved LMXB component is not important in soft X-ray) and the stellar contribution (only the CV/AB contribution) to show the relative significance of the hot gas.}\label{fig:NGC1291RadialProfile}
\end{center}
\end{figure}

\emph{NGC~1291} --- This galaxy contains several local IR features: two central spiral arms and an outer ring (Fig.~\ref{fig:image}a). However, the $24\rm~\mu m$ luminosity of the galaxy is not significantly higher than that expected from the circumstellar dust heated by old evolved stars (Temi et al. 2009), so globally SF is not energetically important, although weak SF may still exist in the outer ring. NGC~1291 is rich in atomic gas (Table~\ref{table:basicpara}), which distributes in a ring-like structure associated with the IR-bright ring (van Driel et al. 1988; Hogg et al. 2001; Irwin et al. 2002). The bump of the radial profile in PAH and $24~\rm\mu m$ emissions at $\sim4\farcm5$ shows the position of this IR-bright ring, which is well outside the soft X-ray bump at $\sim3^{\prime}$ (Fig.~\ref{fig:NGC1291RadialProfile}). We thus conclude that no X-ray counterpart can be clearly associated to this IR-bright gas-rich ring. The soft X-ray enhancement inside it is therefore unlikely directly produced by SF activities. Instead, it may be produced by the mass loading of the cold gas in the ring to the hot phase.

\emph{NGC~2681} --- High resolution optical observations revealed multiple bar-like stellar structures while no clear spiral arms in this galaxy (Erwin \& Sparke 1999; Moiseev et al. 2004; also see the zoom-in panel of Fig.~\ref{fig:image}c). Its inner stellar population lacks any measurable color gradient, which indicates that it underwent a starburst $\approx1\rm~Gyr$ ago, encompassing the interior region (Cappellari et al. 2001). A significant amount of molecular gas ($\gtrsim10^7\rm~M_\odot$) is detected in the very central region of this galaxy (within a diameter of $\sim0\farcm25$, Taniguchi et al. 1994), but the amount and structure of the cold gas in the outer disk is still not clear. The X-ray image of NGC~2681 shows two clear arm-like structures (first discovered by Kilgard et al. 2005), which have no corresponding optical or near-IR counterparts. More detailed observations of cold gas in this galaxy may help to reveal the nature of these spiral-arm-like structures.

\emph{NGC~2787} --- NGC~2787 is rich in cold gas as a low mass S0 galaxy, but has a low molecular-to-atomic gas mass ratio, $\sim2\%$, among those with the lowest molecular-to-atomic gas mass ratio in Welch \& Sage (2003, 2006)'s S0 galaxy sample. Most of the atomic gas distributes in a ring-like structure with a diameter of $\sim6\farcm4$ (Shostak 1987), significantly larger than the extent of the stellar light. The unresolved soft X-ray emission shows smooth featureless morphology, the spectrum of which is consistent with primarily stellar in origin, with only a small contribution from hot gas (Table~\ref{table:hotgas}).

\emph{NGC~3115} --- This galaxy is a highly evolved, quiescent system, with a relatively old stellar population (Norris et al. 2006) and little on-going SF as traced by the weak mid-IR emission (Fig.~\ref{fig:image}g). It is also the most gas-poor galaxy in the present sample, with only upper limits in the amount of the molecular and atomic gases (Table~\ref{table:basicpara}). The spectral analysis of the unresolved X-ray emission shows that it mainly consists of unresolved stellar sources, with little contribution from the hot gas
(Table~\ref{table:hotgas}). As a nearly edge-on galaxy, NGC~3115 does not show any extinction features similar to those detected in NGC~5866 (Paper~I), which suggests a very low cold gas content, consistent with CO observations (Welch et al. 2010).

\emph{NGC~5866} --- This is probably the most molecular gas-rich galaxy in our sample (Welch \& Sage 2003; the molecular gas mass of NGC~2681 in the outer disk is not well constrained). As introduced in \S\ref{sec:Introduction}, it hosts a gaseous disk and extraplanar filaments indicating cold gas being blown out from the galactic disk. Analysis of the soft X-ray spectrum indicates that the Fe abundance is significantly supersolar and higher than those in other early-type galaxies (Paper~I; Ji et al. 2009), which means more Type~Ia SN ejecta is observed in the inner bulge of this galaxy.

\section{Discussion}\label{sec:Discussion}

\subsection{Energy Budget}\label{subsec:Budget}

\begin{figure}[!h]
\begin{center}
\epsfig{figure=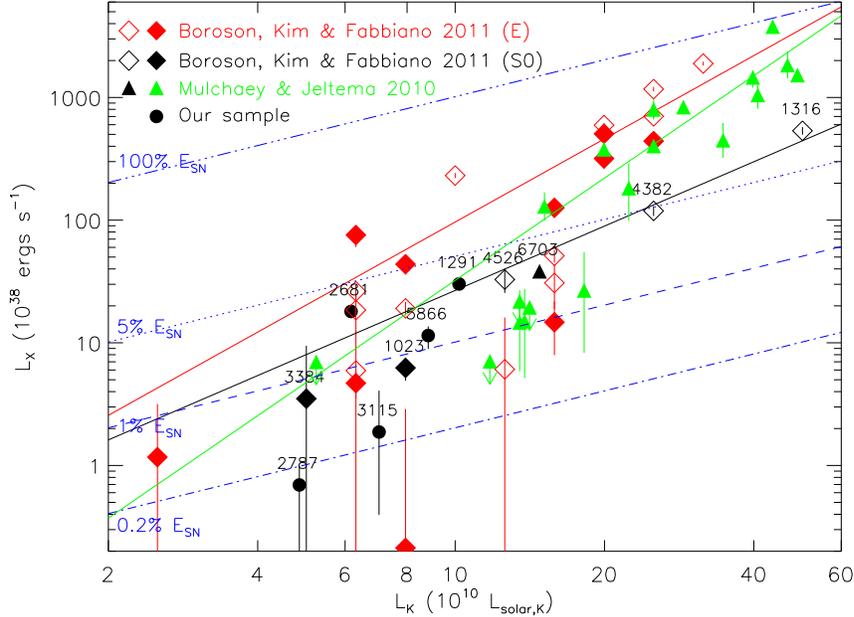,width=0.65\textwidth,angle=0, clip=}
\caption{The soft X-ray luminosity of the thermal component for the hot gas ($L_{X,cor}$ in Table~\ref{table:hotgas}) plotted against the K-band luminosity of the galaxies. The black filled circles are our sample galaxies. The diamonds and triangles are the galaxies in Boroson et al. (2011) and Mulchaey \& Jeltema (2010)'s samples (with CV/AB contribution subtracted), respectively. Filled symbols are galaxies in the field, while open symbols are clustered galaxies. Black symbols are S0 galaxies, while red and green symbols are elliptical galaxies. The NGC names of the S0 galaxies are marked. The black line is the fit to the S0 galaxies in all these three samples. The red and green lines are the fit to the elliptical galaxies in Boroson et al. (2011) and Mulchaey \& Jeltema (2010)'s samples, respectively. The blue lines are the energy input expected from Type~Ia SN feedback.}\label{fig:LK_LX}
\end{center}
\end{figure}

We compare our results with \emph{Chandra} observations of some early-type galaxies from two recent papers (Mulchaey \& Jeltema 2010; Boroson et al. 2011). While only field galaxies are considered in Mulchaey \& Jeltema (2010), the sample in Boroson et al. (2011) includes both field and clustered galaxies, excluding cD ones, the X-ray emission of which may be heavily contaminated by the intra-cluster medium. In Mulchaey \& Jeltema (2010), the CV+AB contribution has not been subtracted. We now subtract it using the same CV/AB model as adopted for our sample (\S\ref{sec:Data}). We further separate the field and clustered galaxies in Boroson et al. (2011)'s sample using the criteria described in \S\ref{sec:SampleSelection}. Using the same criteria, we find galaxies in Mulchaey \& Jeltema (2010) are all in the field, consistent with their own classification. Galaxies in all these three samples are shown in the $L_K-L_X$ plot (Fig.~\ref{fig:LK_LX}).

We estimate the Type~Ia SN energy injection rate, adopting a SN rate of $4.4\times10^{-4}\rm~SN~yr^{-1}/(10^{10}M_\odot)$ (Mannucci et al. 2005), and assuming a mechanical energy input of $10^{51}\rm~ergs$ per SN. Not surprisingly, the observed soft X-ray luminosity is much lower than this energy input, especially for low mass galaxies, consistent with previous results (see Mathews \& Brighenti 2003 and Wang 2010 for recent reviews). For isolated S0 galaxies (black filled symbols), the X-ray luminosity of hot gas typically accounts for $\lesssim5\%$ of the SN heating rate. The ``\emph{missing energy}'' strongly indicates the presence of a galactic wind or subsonic outflow (e.g. Ciotti et al. 1991; Tang et al. 2009a, b).

The measured temperature of the hot gas is also substantially lower than a naive expectation from the Type~Ia SN heating. Adopting a mass loss rate from evolved stars of $\dot{M}\sim0.03\rm~M_\odot~yr^{-1}/(10^{10}M_\odot)$ (Faber \& Gallagher 1976; Peimbert 1993), the expected mean gas temperature would be $\sim2\rm~keV$, about one order of magnitude higher than those measured for the present sample (Table~\ref{table:hotgas}). This discrepancy is apparently a result of multiple effects, including the inhomogeneity of SN heating (Tang et al. 2009a), the mass-loading from cold gas (Tang et al. 2009b; Paper~I), the diversion of mechanical energy to non-thermal forms, and possibly the uncertainty in the specific energy input of the stellar feedback.

We further separate E and S0 galaxies in Fig.~\ref{fig:LK_LX}. Eskridge et al. (1995a,b) found that S0 galaxies are systematically X-ray fainter than elliptical galaxies. This conclusion is based on their different $L_X$ distribution functions obtained from a large \emph{Einstein} sample of 72 elliptical and 74 S0 galaxies. Many theoretical and observational works have been presented to explain this trend. The two most important mechanisms are the partially rotation-supported gravitational potential and the flattened stellar mass distribution (e.g., Ciotti \& Pellegrini 1996; Brighenti \& Mathews 1996; D'Ercole \& Ciotti 1998).

Rotation is thought to be important in reducing the $L_X$ in two different ways, either by reducing the effective gravitational potential or by moving the gas to larger radii. In the sequential wind, outflow and inflow scenario (Ciotti et al. 1991), Ciotti \& Pellegrini (1996) showed that rotation cannot change the state of the gas flow for any fixed galaxy structure, but can only make the existing flow state more stable. For early-type galaxies in wind or outflow states, rotation could reduce $L_X$, but as the dynamical state is not changed, it only has a minor effect. On the other hand, Brighenti \& Mathews (1996) showed that in hydrostatic or inflow states, rotation can significantly reduce the $L_X$ in sense of flattening the X-ray
isophotes. Due to the increasing ordered motion introduced by the rotation, more gas is deposited in a large disk when it cools, and arrives at larger radii after long-term evolution. It can then be removed by ram-pressure stripping or SN feedback. The total effect is to flatten the central diffuse X-ray surface brightness distribution and reduce the observed $L_X$. Different optical and diffuse X-ray morphologies have been evidenced in normal elliptical galaxies, but such differences are suggested to be produced by AGN feedback instead of rotation-supported
gravitational potential (Diehl \& Statler 2007, 2008). However, it is still not clear if rotation is more important in quiescent S0 galaxies, which typically have substantial rotation.

Ciotti \& Pellegrini (1996) showed that even in the presence of massive spherical dark matter halos, the flattening of stellar mass distribution (S0 or elongated elliptical galaxies) is sufficient to change the dynamical state of the gas flow from inflow to wind. Such a flattening will greatly reduce the observed $L_X$. D'Ercole \& Ciotti (1998) further showed in their 2-D simulations that the flattening of the stellar mass distribution can cause decoupling of the gas flows, i.e., after the initial wind phase, the flow tends to revert to an inflow in the polar region, while keep outflowing in the outer disk. This decoupled gas flow results in a lower $L_X$ of flattened galaxies than that of spherical galaxies.

As shown in Fig.~\ref{fig:LK_LX}, high mass S0 galaxies, in particular NGC~4382 and NGC~1316, are indeed X-ray fainter than elliptical galaxies of the same stellar masses. However, at the low mass end, E and S0 galaxies tend to be indistinguishable in $L_X$. We combine all the S0 galaxies in the three samples; by jointly fitting them, we obtain an $L_K-L_X$ relation as:
\begin{equation}\label{equi:LKLXAllS0} \log(L_X/{10^{38}\rm~ergs~s^{-1}})=(-0.3\pm0.4)+(1.7\pm0.3)\log(L_K/10^{10}\rm~L_{\odot,K})
\end{equation}
In comparison, the similar fittings of the elliptical galaxies in Mulchaey \& Jeltema (2010) (Eq.~\ref{equi:LKLXMulchaeyE}) and Boroson et al. (2011)'s (Eq.~\ref{equi:LKLXBorosonE}) samples result in:
\begin{equation}\label{equi:LKLXMulchaeyE} \log(L_X/{10^{38}\rm~ergs~s^{-1}})=(-1.3\pm1.5)+(2.8\pm1.0)\log(L_K/10^{10}\rm~L_{\odot,K})
\end{equation}
\begin{equation}\label{equi:LKLXBorosonE} \log(L_X/{10^{38}\rm~ergs~s^{-1}})=(-0.3\pm0.5)+(2.3\pm0.4)\log(L_K/10^{10}\rm~L_{\odot,K})
\end{equation}
The fitting weights of different data points are inversely proportional to the uncertainties. Boroson et al. (2011) adopts a slightly lower CV+AB contribution than that adopted in the present work, so the $L_X$ of their galaxies should be systematically higher. The difference is only $\sim25\%$ of the CV+AB luminosity, insignificant for most of the galaxies included in the plot, especially the high mass ones. As indicated by the fits (Eqs.~(\ref{equi:LKLXAllS0}-\ref{equi:LKLXBorosonE}) and the solid lines in Fig.~\ref{fig:LK_LX}), S0 galaxies tend to have lower $L_X$ than elliptical galaxies for $L_K\gtrsim10^{11}\rm~L_{\odot,K}$, which is consistent with previous observational and theoretical works as discussed above. However, it is noteworthy that the fits are based on a small sample of 11 S0 galaxies, and only about half of them have $L_K\gtrsim10^{11}\rm~L_{\odot,K}$. Below this luminosity, the scatter of $L_X$ is quite large, E and S0 galaxies are not significantly different.

\subsection{The Cold-Hot Gas Relation}\label{subsec:ColdHotGas}

As introduced in \S\ref{sec:Introduction}, many mechanisms may contribute to the large scatter in the $L_K-L_X$ relation, including the different X-ray properties of E and S0 galaxies as discussed in \S\ref{subsec:Budget}. We now investigate how the cold gas content may affect the X-ray properties of early-type galaxies, and hence the scatter of the $L_K-L_X$ relation.

\begin{figure}[!h]
\begin{center}
\epsfig{figure=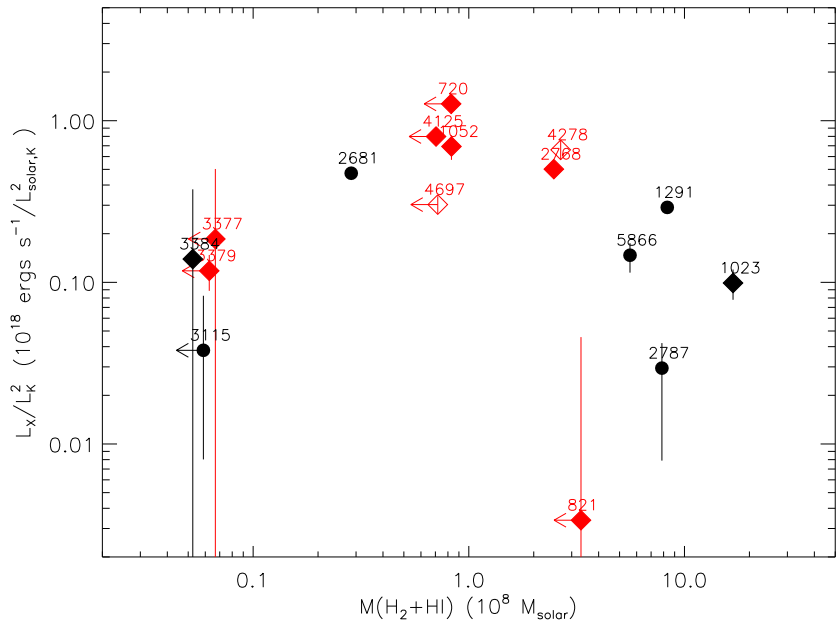,width=0.65\textwidth,angle=0, clip=}
\caption{The residual of the $L_K-L_X$ relation ($L_X/L_K^2$) plotted against the total cold gas mass (molecular plus atomic hydrogen). Symbols are the same as those in Fig.~\ref{fig:LK_LX}.}\label{fig:ColdM_LX}
\end{center}
\end{figure}

As the slope of the $\log L_K-\log L_X$ relation is $\sim2$ (Eqs~\ref{equi:LKLXAllS0}-\ref{equi:LKLXBorosonE}), we compare the total cold gas mass $M_{H_2+HI}$ with $L_X/L_K^2$. We obtain the cold gas masses mostly from a volume limited sample, which excludes members of the Virgo and Fornax Clusters (Welch \& Sage 2003; Sage et al. 2007; Welch et al. 2010). For NGC~2681, we obtain the atomic and molecular gas masses from Haynes et al. (1988) and Taniguchi et al. (1994), respectively. Limited by the small sample size, the large uncertainty in $L_X$, and that only upper limits to the cold gas masses are available for many galaxies, the scatter of the data points in Fig.~\ref{fig:ColdM_LX} may be expected. Nevertheless, $L_X/L_K^2$ appears to correlate positively with $M_{H_2+HI}$ for cold-gas-poor galaxies ($M_{H_2+HI}\lesssim10^8\rm~M_\odot$), whereas an anti-correlation is apparent for cold-gas-rich ones. These (anti-)correlations with the cold gas content can significantly contribute to the large scatter in the $L_K-L_X$ relation.

In early-type galaxies with a fixed gravitational potential, the observed diffuse X-ray luminosity is mainly determined by the specific energy of hot gas (energy per particle) (e.g., Ciotti et al. 1991; Tang et al. 2009a,b). The higher the specific energy, the faster the gas flows out, hence the lower the observed X-ray luminosity. The specific energy is determined by the ratio of the energy and mass input rates ($\dot{E}/\dot{M}$). We thus explain the observed cold-hot gas relation (Fig.~\ref{fig:ColdM_LX}) by assuming that early-type galaxies, especially S0s, are evolved from cold-gas-rich precursors after a major starburst epoch, without further replenishment of external cold gas. In this scenario, the energy input is then mainly from SNe, while the mass input is from stellar mass loss and gas leftover from star formation. Initially, the energy input rate (especially from prompt Type~Ia SNe, after initial core-collapsed SNe) decreases fast after the cease of starburst (e.g., Ciotti et al. 1991; Tang et al. 2009a), while the mass-loading may be reasonably steady or even saturated, via the evaporation of the leftover cold gas. In comparison, the mass input from stellar mass loss may not be as important at this early stage. As a result, the specific energy of the hot gas is expected to decrease with time, hence the increase of the X-ray luminosity with time. After the prompt Type~Ia SN stage, the energy input rate decreases only slowly. As the bulk of the leftover gas has been evaporated (e.g., when $M_{H_2+HI}\lesssim10^8 M_\odot$) and as the mass input from the stellar mass loss also decreases with time, one may expect that the specific energy of the hot gas could begin to increase with time. This specific energy increase may then explain the correlation between $L_X/L_K^2$ and $M_{HI+H_2}$ at the lower $M_{HI+H_2}$ end. In this scenario, after the major starburst epoch, most of the leftover cold gas is gradually removed by the long-lasting Type~Ia SNe feedback, which also helps to quench the SF.

The general trend of Fig.~\ref{fig:ColdM_LX} could be roughly explained with the above scenario. However, the real case may be much more complicated. For example, the spatial distribution of cold gas may affect the efficiency of cold/hot gas interaction and so the observed hot gas luminosity. As the distribution of hot gas concentrates toward the inner bulge in most cases, the galaxies with a more centrally concentrated distribution of cold gas are expected to have a stronger interaction between the cold/hot gases, such as the case of NGC~5866 (Paper~I). On the other hand, if the cold gas mainly distributes in the outer region, as for the case of the \ion{H}{1} ring of NGC~1291, the cold/hot gas interaction then tends to be less significant (Irwin et al. 2002). However, the amount and spatial distribution of the cold gas seem to be dependent on other parameters, such as the origin of the gas (internal or external), the merger and SF history etc. (e.g., Sage \& Welch 2006; Combes et al. 2007; Crocker et al. 2011). These issues are beyond the scope of this paper, but they may contribute to the large scatter of the cold/hot gas relation (Fig.~\ref{fig:ColdM_LX}).

Clearly, more X-ray and radio observations are needed to test the above scenario. In particular, accurate measurement of the hot gas temperature can be very helpful, as it is a direct measurement of the specific energy. Currently the uncertainties in the counting statistics of the data are too large to tightly constrain the temperature (the errors listed in Table~\ref{table:hotgas} are only the fitting error). In addition, measurements of the cold gas with improved sensitivity are also needed.

\section{Summary}\label{sec:Summary}

To study the hot gas properties and cold-hot gas relation in S0 galaxies, we have conducted a careful analysis of the \emph{Chandra} data of five nearby isolated S0 galaxies. We quantitatively subtract various stellar contributions to reveal the truly diffuse hot gas emission, and further compare the results to those of early-type galaxies from relevant recent papers.

The isolated S0 galaxies studied in the present work are all X-ray faint, with the soft X-ray luminosity of the hot gas typically accounting for $\lesssim5\%$ of the Type~Ia SN energy injection rate. Such a low cooling rate indicates the presence of a galactic superwind or subsonic outflow. The hot gas luminosity of a massive S0 galaxy tends to be lower than that of an elliptical galaxy of the same stellar mass, consistent with previous observational and theoretical works. However, at the low mass end, this difference is not significant.

$L_X/L_K^2$, which roughly describes the residual of the $L_K-L_X$ relation (the slope of the $\log L_K-\log L_X$ relation is $\sim2$ from the analysis in this work) or the relative richness of the hot gas in early-type galaxies, tends to correlate positively with the total cold gas mass ($M_{H_2+HI}$) for cold-gas-poor galaxies with $M_{H_2+HI}\lesssim10^8\rm~M_\odot$, while they anti-correlate with each other for cold-gas-rich galaxies. This cold-hot gas relation, together with the different hot gas properties for E and S0 galaxies, may contribute to the large scatter of the $L_K-L_X$ relation.

We further speculate a scenario to explain the overall cold-hot gas relation, assuming an early-type galaxy with a significant amount of leftover gas from a star forming precursor evolves without replenishment of external cold gas. Initially, the observed X-ray luminosity increases with time primarily due to the decreasing SN energy input rate and hot gas specific energy after the major starburst epoch. This stage corresponds to a relatively cold-gas-rich phase of the galaxies or the anti-correlation part of the cold-hot gas relation. In the late evolutionary stage, however, the X-ray luminosity increases with time due to the decreasing mass input rate from the mass loading of the cold gas into the hot phase. This stage corresponds to the cold-gas-poor galaxies or the positive-correlation part of the cold-hot gas relation. This scenario suggests the leftover cold gas in an early-type galaxy is mainly removed by the long-lasting Type~Ia SN feedback after its precursor's starburst stage.

\acknowledgements

This work is supported by NASA through the CXC/SAO grants AR7-8016A and G08-9088B, by NSFC through the grants 10725312 and 10673003 and the China 973 Program grant 2009CB824800.

\end{document}